\documentclass[twocolumn,showpacs,preprintnumbers,amsmath,amssymb,nofootinbib]{revtex4-1}

\pdfoutput=1
\usepackage{epsfig}
\usepackage{cancel}

\usepackage{amsmath, amsthm, amsfonts, amssymb, latexsym}
\usepackage{upgreek}
\usepackage[caption=false]{subfig}

\usepackage{url}
\usepackage{ifpdf}

\usepackage{dcolumn}
\usepackage{bm}

\def\bequ{\begin{equation}}
\def\eequ{\end{equation}}
\def\be{\begin{equation}}
\def\ee{\end{equation}}

\begin{document}

\title{Marginal scalar and Proca clouds \\
around Reissner-Nordstr\"om black holes}

\author{Marco O. P. Sampaio}
 \email{msampaio@ua.pt}
\author{Carlos Herdeiro}
\email{herdeiro@ua.pt}
 \author{Mengjie Wang}
\email{mengjie.wang@ua.pt}
\affiliation{\vspace{2mm}Departamento de F\'\i sica da Universidade de Aveiro and I3N \\ 
Campus de Santiago, 3810-183 Aveiro, Portugal \vspace{1mm}}%

\date{June 2014}

\begin{abstract}
Massive scalar test fields around Kerr black holes can form quasi-bound states with complex frequencies. Some of these states decay in time, but some others -- in the superradiant regime -- grow, causing a superradiant instability. Precisely at the threshold between decaying and growing modes, there exist bound states with real frequency, known as \textit{scalar clouds}. Fully non-linear counterparts of these clouds have been shown to yield Kerr black holes with scalar hair. Here, we consider massive, (electrically) charged, test scalar and Proca fields on the background of Reissner-Nordstr\"om black holes. By computing the quasi-bound state frequencies, we establish that no such states exist in the superradiant regime for the Proca field -- a similar behavior to that known for scalar fields. But decaying quasi-bound states with an arbitrarily small imaginary part of the frequency exist and thus which are arbitrarily long lived. In the limit of vanishing imaginary part of the frequency, the field does not trivialize and we dub the corresponding configurations as \textit{marginal (charged) scalar or Proca clouds}, since they are only marginally bound. We comment on the possible existence of non-linear counterparts to these marginal clouds.
\end{abstract}


\maketitle

\section{Introduction}
Scalar clouds~\cite{Hod:2012px,Hod:2013zza,Herdeiro:2014goa} are equilibrium configurations of a complex, massive test scalar field in the background of a Kerr black hole (BH). They are stationary bound states, with a real frequency, in the same sense as atom orbitals in elementary quantum mechanics.
These configurations are possible due to the existence of two qualitatively different types of \textit{quasi}-bound states -- i.e. bound field configurations with a \textit{complex} frequency:
\begin{itemize}
\item[i)] Time decaying quasi-bound states; this is the generic behavior expected for matter around a BH, due to the purely ingoing boundary condition at the horizon. Indeed, this is the only kind of quasi-bound state that can be found around Schwarzschild BHs, even if some of these states can be very long lived~\cite{Barranco:2012qs}. 
\item[ii)] Time growing quasi-bound states; this occurs for Kerr BHs in the superradiant regime~\cite{Zeldovich1,Zeldovich2,Press:1972zz,Starobinsky:1973a,Damour:1976kh,Zouros:1979iw,Detweiler:1980uk,Cardoso:2004nk,Cardoso:2004hs,Dolan:2007mj,Hod:2009cp,Rosa:2009ei,Yoshino:2012kn,Witek:2012tr,Dolan:2012yt,Yoshino:2013ofa,Cardoso:2013krh,East:2013mfa,Shlapentokh-Rothman:2013ysa}, i.e. when the real part of the frequency, $\omega$, of the quasi-bound state obeys $\omega<m\Omega_H$, where $m$ is the azimuthal harmonic index of the field mode and $\Omega_H$ is the horizon angular velocity of the Kerr BH. This yields an instability of Kerr BHs in the presence of any field for which quasi-bound states can be found in the superradiant regime. Typical studies use scalar fields but, for instance, an interesting study with the Proca fields has been reported in~\cite{Pani:2012vp}.
\end{itemize}
Scalar clouds exist at the boundary between these two regimes, i.e. when $\omega=m\Omega_H$, which is compatible with the bound state condition $\omega<\mu$ for test scalar fields on the Kerr background. Renewed interest concerning this clouds has been triggered by the recent observation that there are Kerr BHs with scalar hair~\cite{Herdeiro:2014goa,Herdeiro:2014ima,Herdeiro:2014jaa}, found as exact solutions of the Einstein-Klein-Gordon system, and which correspond to the non-linear realization of these scalar clouds.

It has long been known that charged, i.e. Reissner-Nordstr\"om (RN) BHs can amplify charged scalar fields through superradiant scattering~\cite{Bekenstein:1973ur}. This is a process that has qualitative similarities with the superradiant scattering of (neutral) scalar fields by Kerr BHs. For the charged case, however, it is not possible to find quasi-bound states in the superradiant regime. In other words, there is no analogue of a charged superradiant instability for asymptotically flat charged BHs~\cite{Furuhashi:2004jk,Hod:2013nn,Hod:2013eea}.\footnote{Such instability can, however, be achieved by changing the asymptotics of the problem, for instance, imposing a mirror boundary condition~\cite{Herdeiro:2013pia,Hod:2013fvl,Degollado:2013bha,Li:2014gfg} or considering AdS asymptotics~\cite{Wang:2014eha}.} As such, one concludes that there will be no charged scalar clouds around RN BHs analogue to the ones discussed above for Kerr, i.e. as true bound states. An open question is whether the same holds for higher spin massive fields, where an effective potential analysis (which indicates if there is a potential well to support quasi-bound states) is not always straightforward. Here we study the charged Proca field around a RN BH, and show that similar statements to those for a scalar field hold.

Although no stationary bound states exist for charged scalar or Proca fields around RN BHs, we show in this paper that stationary \textit{marginally bound states} do exist. The first observation of such states was reported in~\cite{Degollado:2013eqa} in the double extremal limit wherein the BH charge tends to its mass $|Q|\rightarrow M$ and the field charge tends to its mass $|q|\rightarrow \mu$. It was noticed that, in this limit, the imaginary part of the quasi-bound states vanishes and the real part tends to the mass of the field. Moreover,  in this limit, the scalar field does not trivialize and it can be interpreted as a collection of scalar particles at rest, outside the horizon, in a no-force configuration with the BH, due to a balance of electromagnetic and gravitational forces. These particles are only marginally bound to the BH and we shall therefore dub the corresponding field configuration as \textit{marginal (charged) clouds}. 

In fact we show that one needs not take a double extremal limit (as in~\cite{Degollado:2013eqa}) to get marginal (charged) scalar clouds  around RN BHs. It is enough that the \textit{threshold condition}
 \begin{equation}
qQ=\mu M \ ,
\label{tc}
\end{equation}
is obeyed, clearly indicating that the equilibrium is due to a force balance.\footnote{This fact has been independently realized by J. C. Degollado.} Moreover, we show that a completely analogous behavior is found for a charged Proca field around RN BHs. It is worth emphasizing that although these clouds are only marginally bound, arbitrarily long lived quasi-bound states exist close to the limit where these clouds appear, which, for many practical purposes may be faced as eternal clouds.

The structure of this paper is organized as follows: In Sect.~\ref{sec:qProca} we review the background geometry as well as the field equations for the charged Proca field and summarize the equations for the charged massive scalar field. In particular, in Sect.~\ref{sec:eff_potential}, we analyze the effective potential for a decoupled mode of the Proca field, as to illustrate some expected features of the result. In Sect.~\ref{sec:num_strategy} we describe the numerical strategy to obtain the quasi-bound state frequencies for coupled (in the Proca case) and decoupled modes, summarize some known results in the small frequency limit and combine them to predict an approximation for the (hydrogen-like) real part of spectrum for the charged Proca field. In Sect.~\ref{results} we present numerical results illustrating the variation of the quasi-bound state frequencies with particle charge and mass as to identify the condition for long lived states. In Sect.~\ref{sec_analytic} we demonstrate that there are indeed solutions when the threshold condition holds, which means that the imaginary part of the frequency vanishes. In a double extremal limit we even find closed form solutions for both the scalar (already discussed in~\cite{Degollado:2013eqa}) and also the Proca case. Finally we conclude with a summary and discussion of our results in Sect.~\ref{Discussion}; in particular we discuss the possible existence of non-linear realizations of these marginal clouds.

Throughout this paper we shall use units with $G=1=c=\hbar$.

\section{$U(1)$ charged Proca and scalar fields on a charged black hole}
\label{sec:qProca}

\subsection{The Proca field}

We consider first a Proca field $W^{\mu}$ with mass $\mu$, which is charged under a $U(1)$ gauge symmetry associated with the gauge field $A_\mu$.  The field $W^{\mu}$ can describe $W$ particles in the Standard Model of particle physics (SM) coupled to gravity and its Lagrangian density is
\begin{equation}
\mathcal{L}= -\dfrac{1}{2}W^\dagger_{\mu\nu}W^{\mu\nu}-\mu^2W_\mu^\dagger W^\mu-iqW_\mu^\dagger W_\nu F^{\mu\nu} \; ,\label{Procalag}
\end{equation}
where $W_{\mu\nu}=\mathcal{D}_\mu W_\nu-\mathcal{D}_\nu W_\mu$, $\mathcal{D}_\mu\equiv \partial_\mu-i q A_\mu$. Here $q$ is the charge of the Proca field, and its coupling to the electromagnetic field strength tensor $F_{\mu\nu}=\partial_\mu A_\nu-\partial_\nu A_\mu$ is a consequence of gauge invariance. The background geometry for a $U(1)$ charged BH, is given by the RN solution with line element
\begin{equation}
ds^2=-V(r)dt^2+\dfrac{1}{V(r)}dr^2+r^2(d\theta^2+\sin^2\theta d\varphi^2) \ ,
\end{equation}
where
\begin{equation}
V(r)=1-\dfrac{2M}{r}+\dfrac{Q^2}{r^2}\;.\nonumber
\end{equation}
$M$ and $Q$ are the BH mass and charge parameters, respectively. For numerical convenience, we change to units such that the outer horizon radius is at $r_H=1$, i.e. 
\begin{equation}
2M=1+Q^2 \ .
\end{equation}
The field equations for a Proca field on an Einstein-symmetric (and in particular spherically symmetric) background, were first fully separated into a radial system by us in~\cite{HSW:2012}  to study Hawking radiation in brane world scenarios. In~\cite{HSW:2012} we have shown that the radial wave equations governing these field perturbations, for modes of frequency $\omega$, are composed by a system of two coupled modes $(\psi,\chi)$ obeying
\begin{eqnarray}
&&\left[V^2\dfrac{d}{dr}\left(r^2\dfrac{d}{dr}\right)+\left(\omega r -qQ\right)^2-V\left(\ell(\ell+1)+\mu^2r^2\right)\right]\psi\nonumber  \\ 
&&+\left[2iV\omega r-ir\left(\omega r-qQ\right)V'\right]\chi=0 \ ; \nonumber \\
&&\left[V^2r^2\dfrac{d^2}{dr^2}+\left(\omega r -qQ\right)^2-V\left(\ell(\ell+1)+\mu^2r^2\right)\right]\chi \nonumber \\ && +\left[2iqQV-i r\left(\omega r-qQ\right)V'\right]\psi=0 \; \;, \; \; \; \; \label{coupledequations}
\end{eqnarray}
and a decoupled transverse mode $\Upsilon$
\begin{eqnarray}
&&\left[r^2V\dfrac{d}{dr}\left(V\dfrac{d}{dr}\right)+\left(\omega r  -qQ\right)^2-\left(\ell(\ell+1)+\mu^2r^2\right)V\right] \Upsilon \nonumber \\
&&=0  \label{TmodeEq}\ ,
\end{eqnarray}
where $\ell=1,2,\ldots$. There is also an exceptional mode $\ell=0$ which is described by a decoupled radial equation
\begin{eqnarray}
&&\left[\dfrac{d}{dr}\left(\dfrac{Vr^4}{(\omega r-qQ)^2-Vr^2\mu^2}\dfrac{d}{dr}\right)+\dfrac{r^2}{V} \right. \nonumber \\ && \left. -\dfrac{qQr^2\left((\omega r-qQ) r V'-2qQV\right)}{\left[(\omega r-qQ)^2-Vr^2\mu^2\right]^2}\right]\psi^{(0)}=0 \ .
\end{eqnarray}
The asymptotics of the solutions of such systems has been detailed in~\cite{HSW:2012,Wang:2012tk}. Let us denote a generic field mode (coupled or not to others) by $\psi_i$. Then the near horizon asymptotics for all modes takes the form
\begin{equation}\label{Eq:NH_expansion}
\psi_i= (r-1)^{\mp i\frac{\omega-qQ}{1-Q^2}}\sum_{n=0}^{+\infty} a_i^{(n)} (r-1)^n\ ,
\end{equation}
where the minus sign corresponds to an ingoing boundary condition at the horizon. On the other hand, in the asymptotic region $r\rightarrow +\infty$ the expansion is
\begin{equation}
\psi_i= e^{ i \Phi}\sum_{n=0}^{+\infty}\dfrac{c_{i,+}^{(n)}}{r^n}+ e^{-i \Phi} \sum_{n=0}^{+\infty}\dfrac{c_{i,-}^{(n)}}{r^n}\label{eq:FFexpansion}
\end{equation}
with
\begin{eqnarray}
 \Phi &\equiv& k r+\varphi \log r\nonumber \\ 
 \varphi&\equiv& \dfrac{(\omega^2+k^2)(1+Q^2)-2qQ\omega}{2k} \\
 k&\equiv& \sqrt{\omega^2-\mu^2} \; .
\end{eqnarray}
All the necessary expansion coefficients were provided in~\cite{HSW:2012,Wang:2012tk}. As shown originally by Press and Teukolsky~\cite{Press:1973zz}, if we consider initial data given on a compact support, the late time dynamics of generic linear perturbation is governed by a superposition of a discrete set of solutions in Fourier space. After imposing an ingoing boundary condition a the horizon, there are basically two types of solutions depending on the boundary conditions when $r\rightarrow +\infty$ (considering a mode oscillating as $e^{i\Phi}$):
\begin{itemize}
\item $\Im(k)<0$: these are the  quasi-normal modes which describe damped oscillations which are free to escape the BH potential and asymptotically grow exponentially. 
\item $\Im(k)>0$, these are quasi-bound states, i.e. they describe field configurations which are confined in the outside region of the BH and decay exponentially when $r\rightarrow+\infty$. They are possible if there is a confining potential well where the field can accumulate. The boundary condition for these states can thus be recast as
\begin{equation}
\lim_{r\rightarrow+\infty}\psi_i=0\;.
\end{equation}
 
\end{itemize}
Quasi-bound states are very interesting in the presence of superradiance, since they provide the possibility of an instability~\cite{Press:1973zz,Zouros:1979iw} as discussed in the introduction. If they exist within the superradiant regime, the field is able to extract energy from the BH which accumulates in the confining potential. This would be signaled by an exponential growth of the wave amplitude. The condition for the instability to appear is $\omega_I\equiv\Im{(\omega)}>0$, i.e. the time-dependence is
\begin{equation}
\psi_i\sim e^{-i\omega t}=e^{-i\omega_Rt+\omega_It} \; .
\end{equation} 
In the absence of instabilities (i.e. $\omega_I<0$), the wave amplitude decays exponentially with a lifetime
\begin{equation}
\tau\equiv |\omega_I|^{-1}\; .
\end{equation}
If $\omega_I=0$ is possible, then the state can be truly bound (or marginally bound if $\omega_R=\mu$). If $\tau$ is very large, then the state may be effectively considered bound (rather than quasi-bound) for many practical purposes.

\subsection{Effective potential for the transverse mode}
\label{sec:eff_potential}
\begin{figure}
\begin{center}
\includegraphics[clip=true,trim = 40 450 50 50, width=0.44\textwidth]{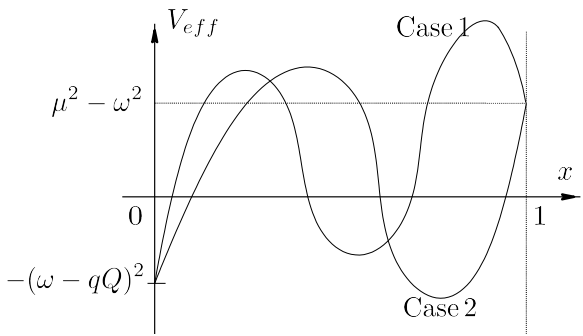}
\end{center}
\caption{\label{EffectiveVschematics} Schematic representation of the effective potential (in the compactified coordinate $x$) for the two possible cases where a well may form. 
}
\end{figure}

A natural strategy to investigate the possibility of quasi-bound states to appear is to recast the radial equations for the fields in a Schr\"odinger like form with an effective potential. In the case of the Proca field, this is not so straightforward for the coupled system. However, the transverse mode Eq.~\eqref{TmodeEq}, can be easily recast in a Schr\"odinger like form ($dr_\star \equiv dr/V $)
\begin{eqnarray}
&&\left[-\dfrac{d^2}{dr_\star^2}+V_{eff}\right] \Upsilon=0\\
&&V_{eff}=\left(\tfrac{\ell(\ell+1)}{r^2}+\mu^2\right)V-\left(\omega -\tfrac{qQ}{r}\right)^2 \; .
\end{eqnarray}
To classify the effective potential and investigate when a well forms, it is convenient to define a compactified coordinate $x\equiv 1-1/r\in [0,1]$ such that 
\begin{equation}\label{eq:Veff}
V_{eff}(x) = \sum_{k=0}^4 b_k x^k \ ,
\end{equation} 
where the coefficients $b_k$ are
\begin{eqnarray}
b_0&=&-(\omega-qQ)^2 \nonumber\\
b_1&=& (1-Q^2)(\mu^2+\ell(\ell+1))-2qQ(\omega-qQ)\nonumber\\
b_2&=& -(2-3Q^2)\ell(\ell+1)+Q^2(\mu^2-q^2)\label{eq:bcoeffs}\\
b_3&=& \ell(\ell+1)(1-3Q^2)\nonumber\\
b_4&=& \ell(\ell+1)Q^2\; . \nonumber
\end{eqnarray} 
Thus we are dealing with a quartic polynomial. Taking into account its values at the end points a well can only form for two possible configurations with three roots  (shown schematically in Fig.~\ref{EffectiveVschematics}).
In case~1 the derivative $V'_{eff}(x)$ must have three positive roots $x_1,x_2,x_3$. Then its form is $V'_{eff}(x)=A(x-x_1)(x-x_2)(x-x_3)$. But we also know that
\begin{eqnarray}
  V'_{eff}(x)&=&\sum_{k=0}^3 (k+1) b_{k+1} x^k\ .
\end{eqnarray}
Equating the two forms we conclude that $x_1x_2x_3=-b_1/(4b_4)$. We are interested in finding a well in the superradiant regime, $\omega<qQ$; but using Eqs.~\eqref{eq:bcoeffs} we conclude that $-b_1/b_4<0$, which is inconsistent with the positivity of the roots of the derivative. Thus case~1 is not possible in the superradiant regime. 

Case~2 on the other hand is defined by\footnote{In this case we could not extract any other condition as for case~1.}
\begin{eqnarray}
V'_{eff}(1)>0 &\Leftrightarrow& \mu^2 >\dfrac{2qQ\omega}{1+Q^2}\\
&\Leftrightarrow& \mu M >qQ\dfrac{\omega}{\mu}
\end{eqnarray} 
which in principle is compatible with the superradiance condition. However, we have made a numerical scan of the parameter space to try to find combinations of parameters satisfying both conditions and found none. 

This analysis indicates that no potential well is possible in the superradiant regime for the transverse modes. For the coupled system it is not simple to make a similar analysis, so we proceed in the following sections investigating the quasi-bound state frequencies numerically and attempting to flow the parameters into the superradiant regime. 

We should note at this point that Furuhashi and Nambu~\cite{Furuhashi:2004jk} found a condition for bound states to exist for a charged scalar field, working in a small frequency and charge approximation, and using an analytic matching technique. They found the condition that $\mu M\gtrsim qQ$, which, from the applicability of their analysis, is not necessarily true for large parameters. We will find, however, a perfect agreement with this threshold condition for large parameters as well, therefore away from the applicability of the analytic matching approximation in~\cite{Furuhashi:2004jk}. 
 
\subsection{The scalar field}
\label{sec:scalar}
In this study we have also solved the problem for a charged massive scalar field. In addition to being interesting on its own this also serves as a comparison to check some of the features we have found for the Proca field. 

Discussions of the charged massive scalar field can be found, for example, in~\cite{Sampaio:2009ra,Sampaio:2009tp,Marcothesis,Degollado:2013eqa} and in~\cite{Furuhashi:2004jk} where the rotating case was also analyzed. Here we only present the potential since the structure is very similar to the $\Upsilon$ mode of the Proca field. After decomposing the scalar field $\varphi$ in spherical harmonics $\varphi=e^{-i\omega t}Y_\ell^m(\phi,\theta)R(r)/r$, the radial equation takes exactly the same Schr\"odinger like form as for $\Upsilon$ with a modified effective potential
\begin{equation}
V_{eff}^{(scalar)}=\left(\tfrac{\ell(\ell+1)}{r^2}+\mu^2+\dfrac{1}{r}\dfrac{dV}{dr}\right)V-\left(\omega -\tfrac{qQ}{r}\right)^2 \; .
\end{equation}
Then all the asymptotic analysis and boundary conditions follow our previous discussion, Eqs.~\eqref{Eq:NH_expansion} and~\eqref{eq:FFexpansion}, for the Proca field modes.

\section{Numerical strategy}\label{sec:num_strategy}
To find the quasi-bound state frequencies, we scan over $\omega$ on the complex plane. We first define a function of $\omega$ which returns the value of the solutions $\psi_i$ at a large $r_{far} \sim (50\sim100)/k_R$ ($k_R=\Re(k)$), i.e. at a very large distance in multiples of the typical wavelength of the wave. For each value of $\omega$ this is done by integrating the radial equations outwards from the horizon, with initial conditions given by the power series expansions Eq.~\eqref{Eq:NH_expansion} evaluated at $r=1.001$, with typically twenty terms.  This function is expected to be exponentially large for generic values of $\omega$ except in the vicinity of a quasi-bound state where it should be very small. Numerically, the quasi-bound state frequency is never attained exactly since there is always some numerical contamination of the exponentially growing solution. For example for a decoupled mode
\begin{equation}\label{eq:illustratedecay}
\psi_i\sim c_{i,+}^{(0)}e^{ik_Rr-|k_I|r}+\ldots+c_{i,-}^{(0)}e^{-ik_Rr+|k_I|r} \; .
\end{equation}   
In the quasi-bound state limit $c_{i,-}^{(0)}\rightarrow 0$. Numerically this corresponds to frequencies that minimize $|\psi_i|_{r_{far}}$. 
\begin{figure*}
\begin{center}
\hspace{-3mm}\includegraphics[clip=true,trim = 90 400 50 30, width=0.51\textwidth]{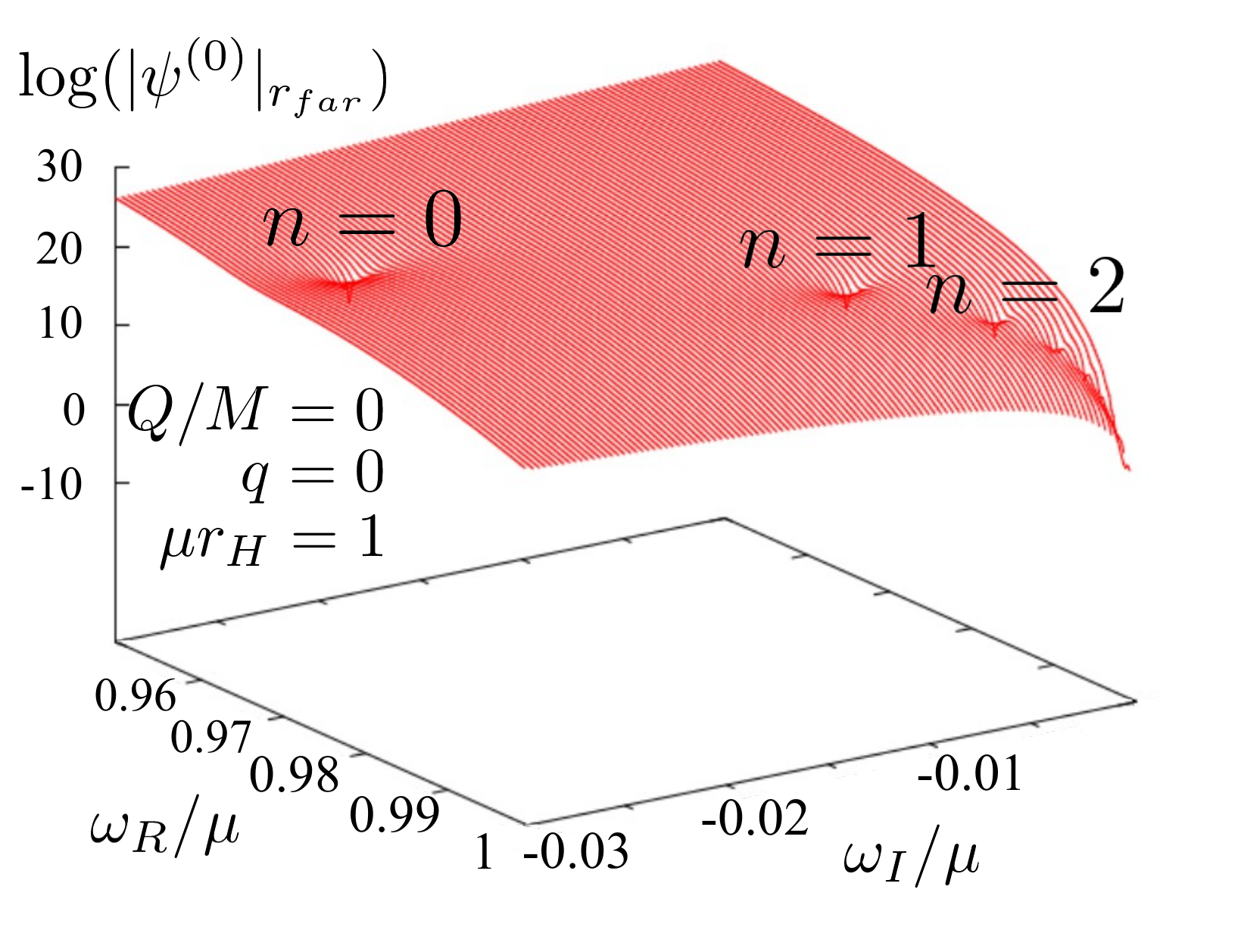}\includegraphics[clip=true,trim = 90 30 50 380, width=0.51\textwidth]{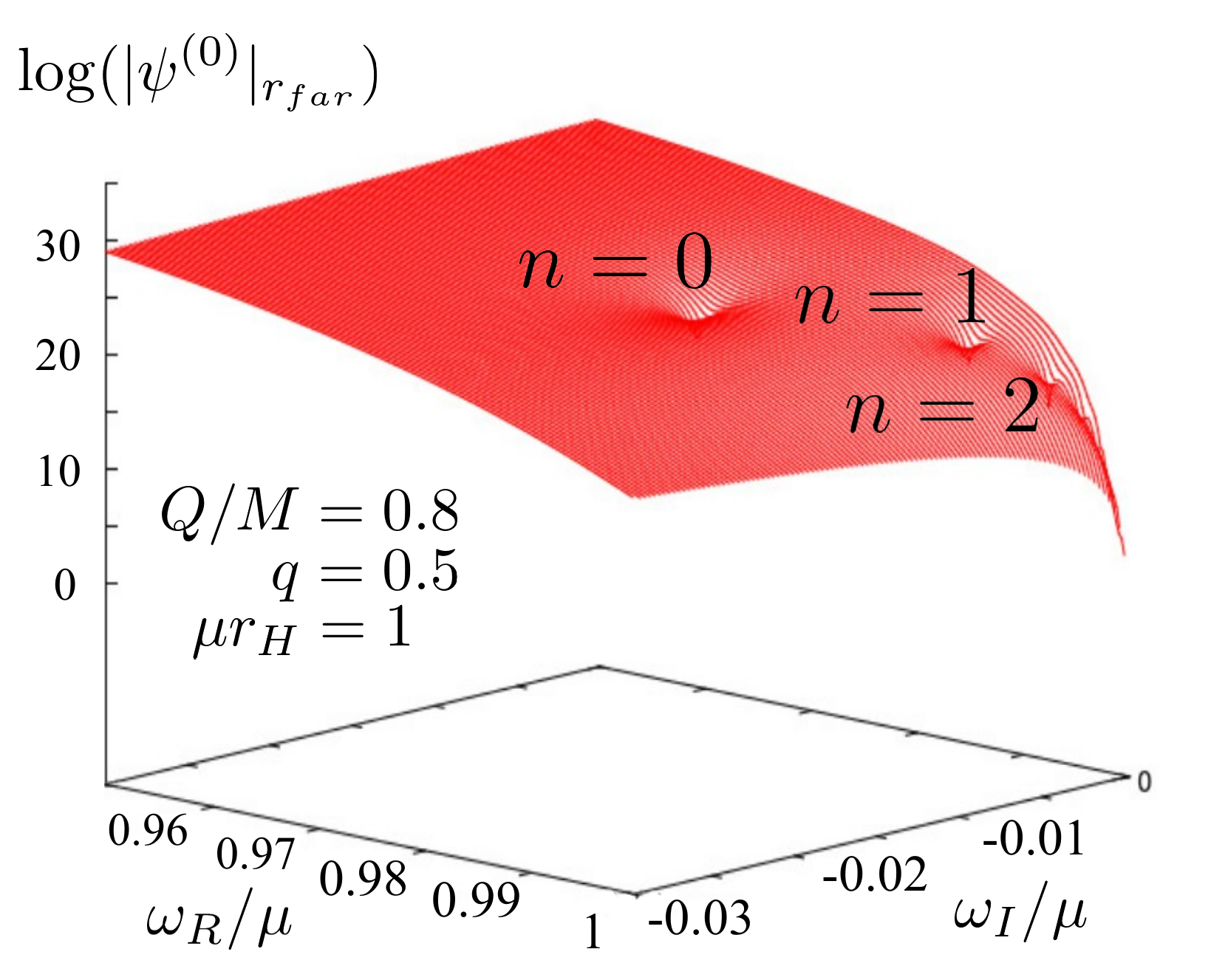}
\end{center}
\caption{\label{SurfacesPsi0} Magnitude of the $\psi^{(0)}$ mode at $r_{far}$ (i.e. in the asymptotic far region), as a function of the complex frequency $\omega$ in the neutral limit (left) and for non-zero charges (right). The quasi-bound state frequencies occur at sharp dips indicated by the level numbers $n$.
}
\end{figure*}
In Fig.~\ref{SurfacesPsi0} we show this quantity for the $\ell=0$ decoupled mode in the neutral case (left) and in the charged case (right). The quasi-bound state frequencies are clearly seen as sharp dips in the surface plots (note that in the vertical axis the logarithm of the scalar field value is represented). We observe that turning on the charge, with $qQ>0$, the real part of the frequency, $\omega_R$, moves to values closer to the field mass, $\mu$, and the (negative) imaginary part $\omega_I$ moves to values closer to zero.

For the coupled sector, the procedure is analogous if one recalls the scattering matrices defined in~\cite{HSW:2012}. First one notes that the general solution of a system of $n$ coupled fields $\psi_i$, with second order linear equations, can be represented by $2n$ integration constants. Those constants can be defined either at the event horizon or at infinity. The linearity of the system implies that a linear transformation relates the integration constants at the horizon, with those at infinity. We denote the ingoing and outgoing wave coefficients at the horizon ($+/-$ respectively)
 \[\vec{\mathbf{h}}=({\mathbf h}^+,{\mathbf h}^-)=(h^+_{i},h^-_{i}) \ , \] where $i=1,2$ for the current coupled system, and, similarly, the coefficients at infinity are defined
  \[\vec{\mathbf{y}}=({\mathbf y}^+,{\mathbf y}^-)=(y^+_{i},y^-_{i}) \ . \]
The linear transformation is then represented as
\begin{equation}
\vec{\mathbf{y}}=\mathbf{S} \vec{\mathbf{h}} \ \ \Leftrightarrow \ \ \left(\begin{array}{c} {\mathbf y}^+ \\ {\mathbf y}^- \end{array} \right)=\left(\begin{array}{c|c} {\mathbf S}^{++} & {\mathbf S}^{+-} \\ \hline {\mathbf S}^{-+}  & {\mathbf S}^{--} \end{array} \right)\left(\begin{array}{c} {\mathbf h}^+ \\ {\mathbf h}^- \end{array} \right)  ,
\end{equation}
where the scattering matrix $\mathbf{S}$ depends on $\omega$, $\ell$, field couplings and the background. It encodes all the information on the scattering process and it can be constructed from specific combinations of modes with boundary conditions set at the horizon.

\begin{figure*}
\begin{center}
\hspace{-3mm}\includegraphics[clip=true,trim = 90 400 50 30, width=0.505\textwidth]{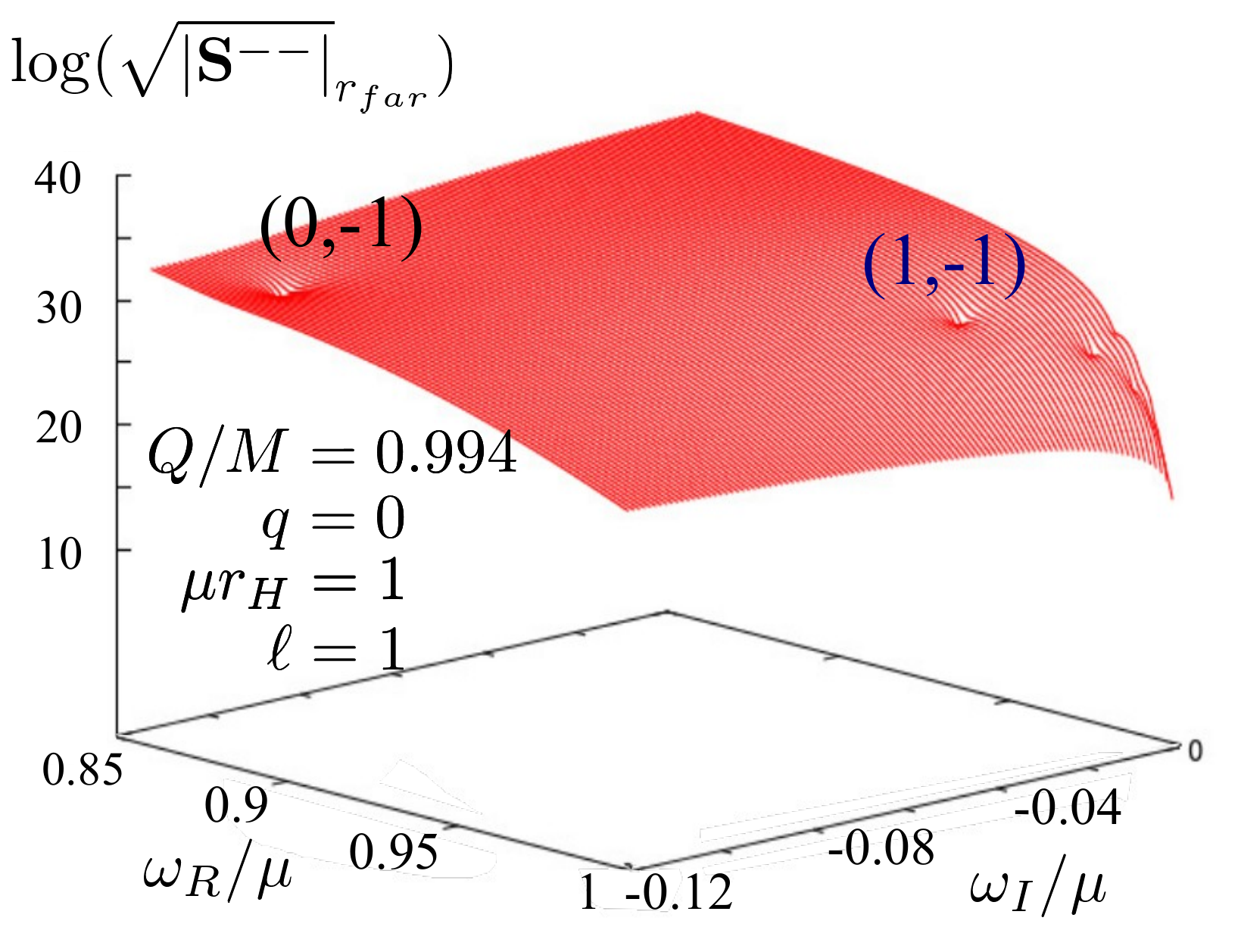}\includegraphics[clip=true,trim = 90 30 50 380, width=0.505\textwidth]{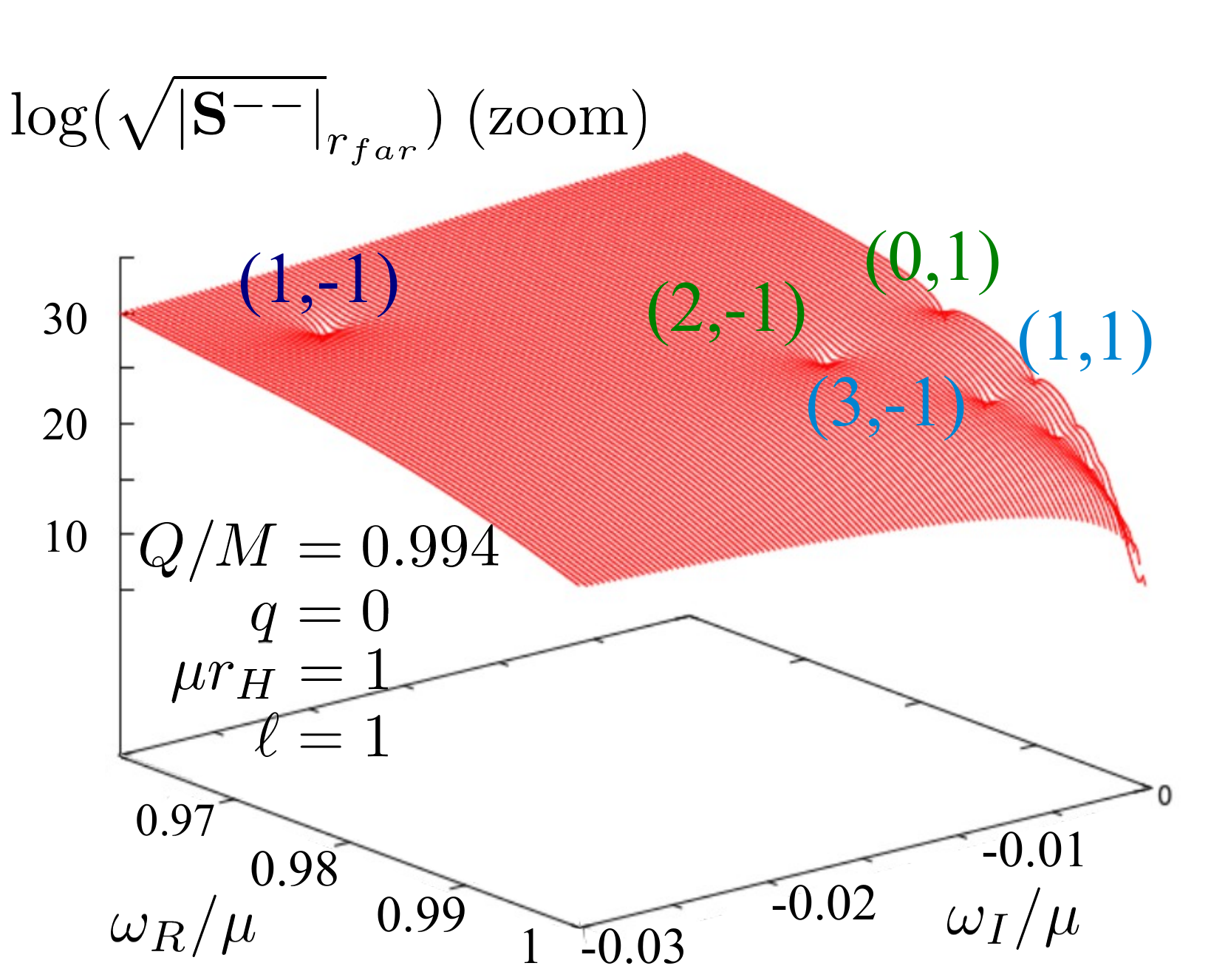}
\end{center}
\caption{\label{SurfacesCoupled} Magnitude of the determinant ${\mathbf S}^{--}$for the coupled system at $r_{far}$ (i.e. in the asymptotic far region), as a function of the complex frequency $\omega$ in the neutral limit for $\ell=1$. The various levels are indexed by $(n,S)$ as discussed in the text. The right panel shows a zoom to better display the positions of a few higher levels.
}
\end{figure*}

Similarly to the decoupled modes, we want to impose an ingoing boundary condition at the horizon i.e. ${\mathbf h}^{+}=0$. Then the solution coefficients far away are
\begin{equation}\label{scattering_ingoing}
{\mathbf y}^s={\mathbf S}^{s-}{\mathbf h}^{-} \; .
\end{equation}
We then must impose that the coefficients of the solution which grows exponentially fast vanish (in analogy to Eq.~\eqref{eq:illustratedecay}), i.e. we need that there is a particular initial condition $\hat{\mathbf h}$ at the horizon such that $\mathbf{y}^{-}=0$, i.e. 
\begin{equation}
0={\mathbf S}^{--}\hat{\mathbf h}^{-} \; .
\end{equation}
Thus we need to choose the eigenvector with zero eigenvalue of ${\mathbf S}^{--}$. The condition for this solution to be possible is then 
\begin{equation}
\det{\mathbf S}^{--}=0 \; ,
\end{equation} 
which will occur at the quasi bound state frequencies. Furthermore, one can check that (up to a normalization constant) the radial profile of the quasi-bound state solution is obtained using the following initial condition at the horizon
\begin{equation}
\left(\begin{array}{c}\hat{h}^{-}_1 \\ \hat{h}^{-}_2\end{array}\right)\propto\left(\begin{array}{c}-S^{--}_{12} \\ S^{--}_{11}\end{array}\right) \; .
\end{equation}
In practice, there is numerical contamination of exponentially growing eigenmodes, which means we must again employ a minimization condition (similarly to the decoupled modes)
\begin{equation}
  \min_{\omega}\left[|{\mathbf S}^{--}|_{r_{far}}\right] \; .
\end{equation}
In Fig.~\ref{SurfacesCoupled} we show an example of this quantity for $\ell=1$ coupled modes with a non-zero background charge and a zoom around the corner where the higher frequency levels pile up (right). We discuss in more detail the effect of the field charge in Sects.~\ref{results} and~\ref{Discussion} so here we only highlight the differences compared with the decoupled cases.
Once again the procedure is to find the local minimum for each valley found in the figure.

Since the search for the quasi-bound state frequencies reduces to a (local) minimization problem, one needs in general a good starting guess which is close enough to the valley. Several analytic estimates have been developed in the literature for small parameters. The general conclusion is that $\omega_R$ follows a hydrogen like spectrum. Of particular relevance to our problem is the charged scalar spectrum approximation found in~\cite{Furuhashi:2004jk}:
\begin{equation}\label{AnalyticFN}
\omega_R\simeq \mu\left[1-\frac{1}{2}\frac{(\mu M-qQ)^2}{N^2}\right]\;,\;N=\ell+n+1
\end{equation}
with $n\in \mathbb{N}_0$.
This approximation suggests that, as the charge is turned on to positive values, there is a critical value at $\mu M =qQ$ after which there is an exit from the quasi-bound state regime (where $\omega_R>\mu$).

Also in the low energy limit, but in the neutral case ($q=Q=0$), a low energy approximation was found for the Proca field in~\cite{Rosa:2011my}. Their expression can be generalized to our case by introducing a charge dependence as an educated guess suggested by the scalar case formula~\eqref{AnalyticFN} i.e.
\begin{equation}\label{eq:ProcaOmegaR}
\omega_R\simeq \mu\left[1-\frac{1}{2}\frac{(\mu M-qQ)^2}{N^2}\right]\; \;, \;  N=\ell+S+n+1
\end{equation} 
where $S=0$ for the transverse modes and $S=1,-1$ for the coupled modes (for the exceptional mode $\ell=0$ there is only $S=1$). Even though in this study we have not developed the analytic matching calculation, we found an excellent agreement (for small parameters) with~\eqref{eq:ProcaOmegaR}.

In Fig.~\ref{SurfacesCoupled} we indicate the labels $(n,S)$ corresponding to this approximation, for the first few pairs of levels which arise in the coupled system. As expected, due to the degeneracy in $N$, i.e the same $N$ can be obtained by adding different combinations of $(n,S)$ for the same $\ell$, the number of degrees of freedom is doubled. This can be observed in the existence of two distinct lines of valleys in the right panel. By continuity, this classification must hold also for larger parameters, so a natural strategy to obtain new frequencies is to perform a flow of the parameters in small steps from some reference quasi-bound state frequencies. This can be pictured as a flow of the valleys in the plots of Figs.~\ref{SurfacesPsi0} and~\ref{SurfacesCoupled}.

Even though analytic approximations give a very useful guide to the dependence of the frequencies, and help understanding the labels of the various levels, in practice (especially for larger parameters) we determined various reference initial estimates for the frequencies of each state graphically (using plots such as Fig.~\ref{SurfacesPsi0}) and then refined them through minimization before using as seeds to the flows.
\begin{table}
\begin{center}
$Qr_H=0$\vspace{1mm}\\
\begin{tabular}{||c|c|c||}
\hline
 $N$ & $(\ell,n,S)$ &$\mu^{-1}(\omega_R,\omega_I)$ \\
\hline
$1$ & $(1,0,-1)$ &$\phantom{..}(0.95972157,-0.0037679893)\phantom{.}$\\
\hline
    & $(1,1,-1)$ & $(0.99032863,-0.00062350043)$\\
$2$ & $(0,0,1)$  & $(0.99031646,-0.00051223870)$\\
    & $(1,0,0)$  & $(0.99131023,-0.00001346974)$\\
\hline
\end{tabular}\vspace{2mm}\\
$Qr_H=0.3 \; (Q/M\simeq 0.55\;,\;\mu M \simeq 0.27)$ \vspace{1mm}\\
\begin{tabular}{||c|c|c||}
\hline
 $N$ & $(\ell,n,S)$ &$\mu^{-1}(\omega_R,\omega_I)$ \\
\hline
$1$ & $(1,0,-1)$ & $(0.95270408,-0.0046419907)$\\
\hline
     & $(1,1,-1)$ & $(0.98838598,-0.00084694923)$\\
 $2$ & $(0,0,1)$  & $(0.98812624,-0.00095739793)$\\
     & $(1,0,0)$  & $(0.98953577,-0.00002375643)$\\
\hline
\end{tabular}\vspace{2mm}\\
$Qr_H=0.9\;(Q/M\simeq 0.994\;,\;\mu M \simeq 0.45)$ \vspace{1mm}\\
\begin{tabular}{||c|c|c||}
\hline
 $N$ & $(\ell,n,S)$ &$\mu^{-1}(\omega_R,\omega_I)$ \\
\hline
$1$ & $(1,0,-1)$ & $\phantom{..}(0.87274751,-0.031112471)\phantom{...}$\\
\hline
    & $(1,1,-1)$ & $(0.96469871,-0.011098631)$\\
$2$ & $(0,0,1)$  & $(0.96824273,-0.016591447)$\\
    & $(1,0,0)$  & $(0.96354486,-0.001705819)$\\
\hline
\end{tabular}
\end{center}
\caption{\label{TabSeeds} Some reference frequencies used in the flow for the Proca field. Note that $\mu r_H=0.5$ and $q=0$ for all the reference points. We indicate the background charge both in inverse horizon radius units (the natural units used throughout this article), and also in BH mass units for comparison.}
\end{table}
In Table~\ref{TabSeeds}, we provide a set of seed frequencies that were used as starting points to produce various of the plots in the results Sect.~\ref{results} for the Proca field. Some scalar frequencies  that were used to obtain scalar profiles to compare with the Proca field in the region of parameters where long lived states were found (see Sects.~\ref{results} and~\ref{Discussion}) are shown in Table~\ref{TabSeedsScalar}.
\begin{table}
\begin{center}
\begin{tabular}{||c|c|c||}
\hline
 $N$ & $(\ell,n)$ &$\mu^{-1}(\omega_R,\omega_I)$ \\
\hline
$1$ & $(0,0)$ & $\phantom{..}(0.93713433,-0.083684629)\phantom{...}$\\
\hline
    & $(0,1)$ & $(0.97694630,-0.016936300)$\\
$2$ & $(1,0)$ & $(0.96780454,-0.000797486)$\\
\hline
\end{tabular}
\end{center}
\caption{\label{TabSeedsScalar} Some reference frequencies used in the flow for the scalar field. Note that $\mu r_H=0.5$ and $q=0$ for all the reference points and that we have focused on $Q/M=0.994$ to compare with the Proca field.}
\end{table}

\section{Results}
\label{results}
In this section, we present a selection of numerical results, focusing mostly on the positive charge coupling case. First, we note that we did not find any quasi-bound states in the superradiant region $\omega<qQ$. 
\begin{figure*}
\begin{center}
\includegraphics[clip=true, width=0.33\textwidth]{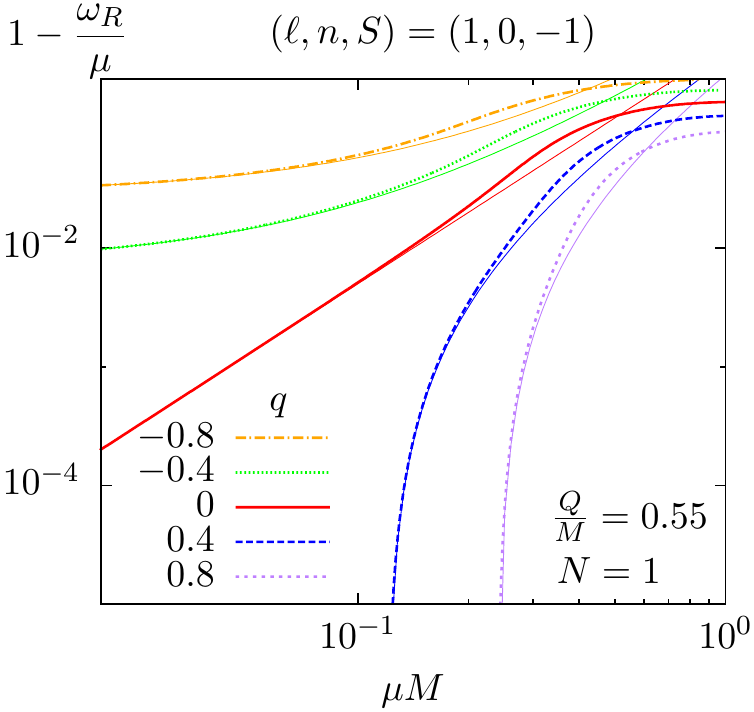}\includegraphics[clip=true, width=0.33\textwidth]{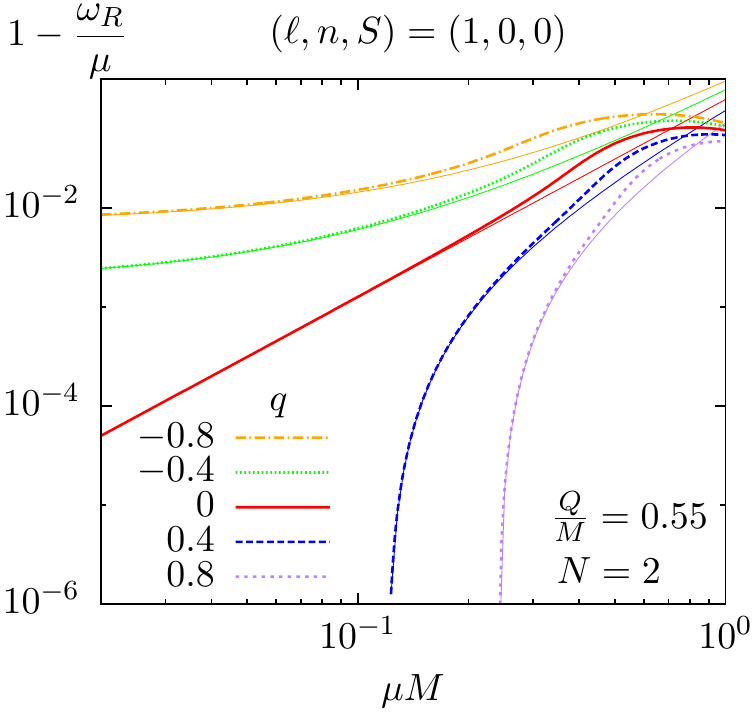}\includegraphics[clip=true, width=0.33\textwidth]{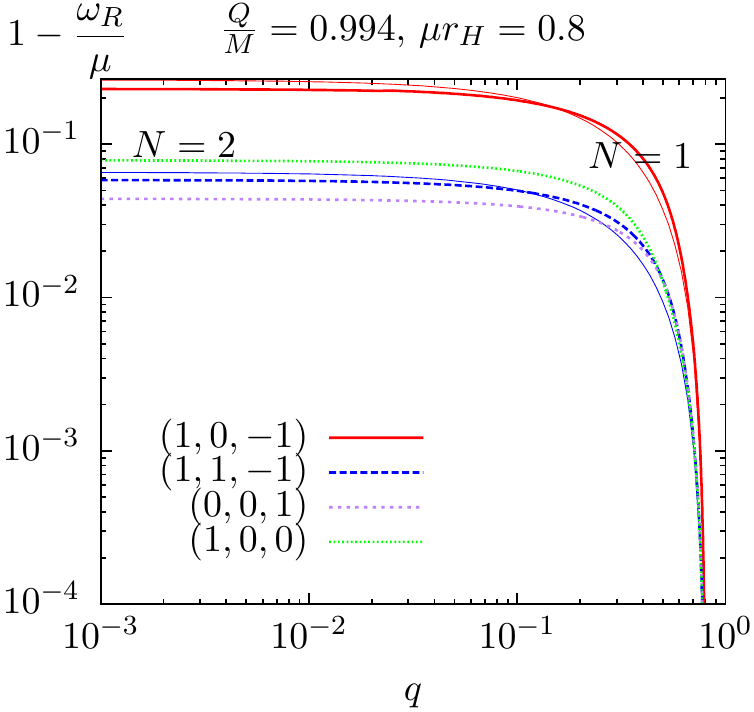} \vspace{2mm}\\
\includegraphics[clip=true, width=0.33\textwidth]{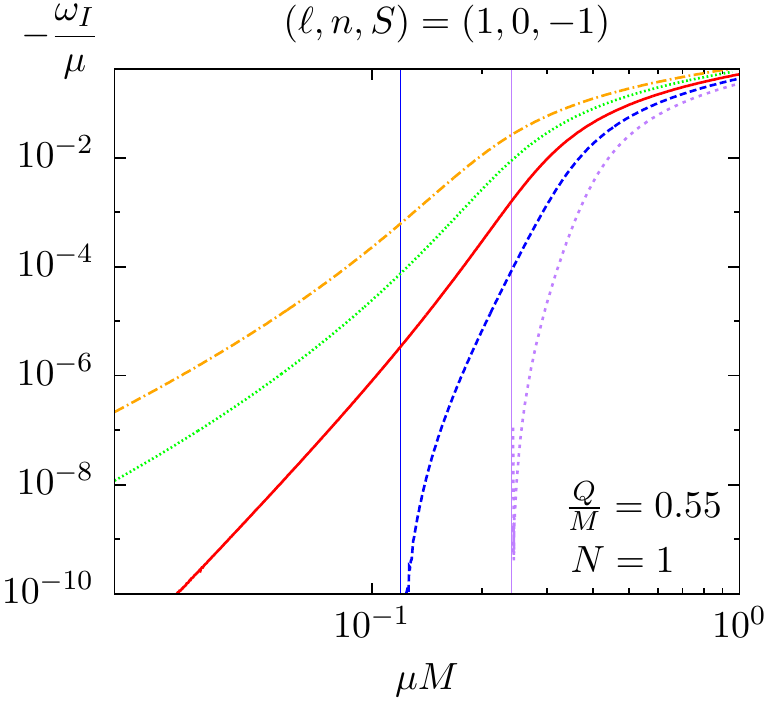}\includegraphics[clip=true, width=0.33\textwidth]{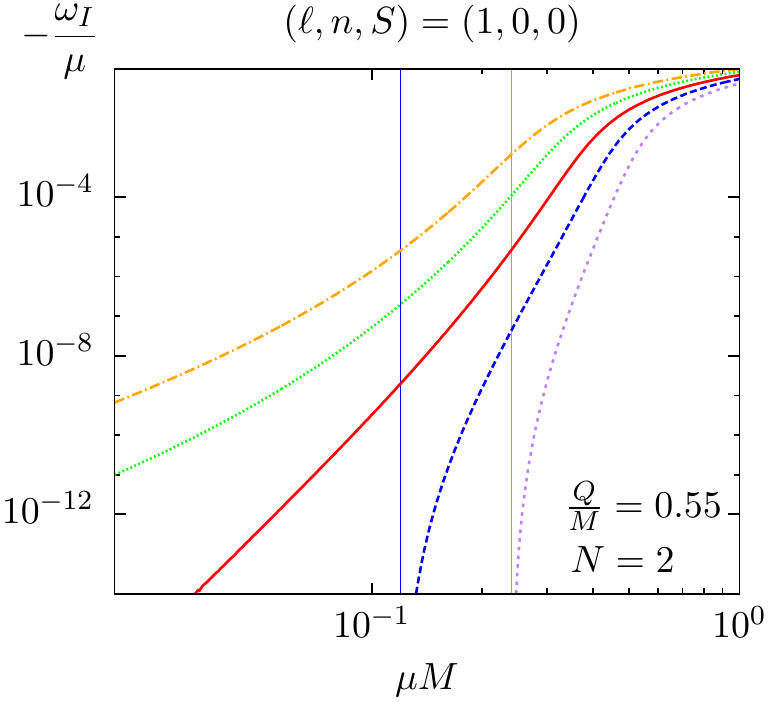}\includegraphics[clip=true, width=0.33\textwidth]{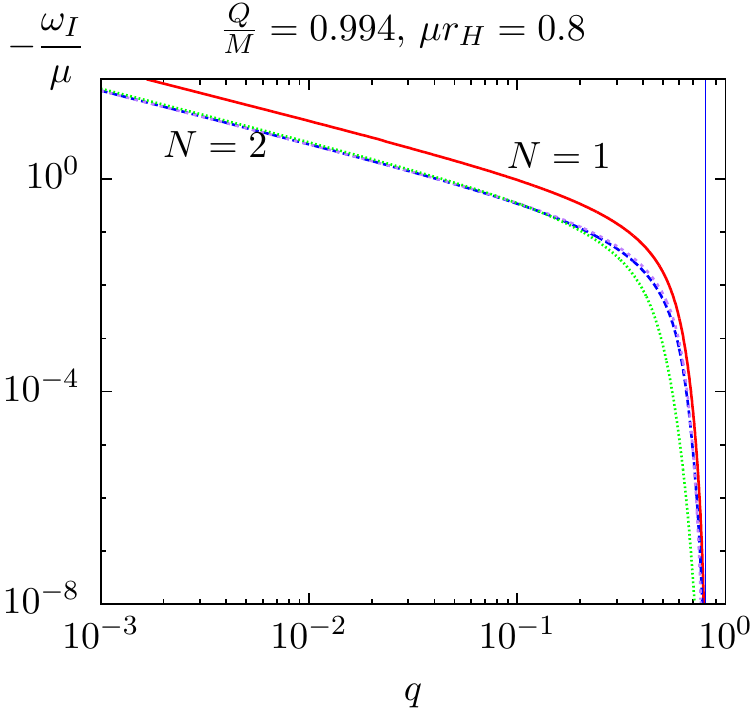}
\end{center}
\caption{\label{Fundamental} Variation of the quasi-bound state frequencies with field mass $\mu$ and charge $q$ for the Proca field. In the top panels we represent curves for the real part $\omega_R$ whereas the corresponding curves for the imaginary part are in the bottom panels. The thick curves labeled in the key to the figures are obtained by numerical integration. The thin solid lines correspond to the analytic formula Eq.~\eqref{eq:ProcaOmegaR}. The vertical lines give, for each case $\mu M=qQ$. 
}
\end{figure*}
To understand why this is the case we analyze first Fig.~\ref{Fundamental} where we represent the variation of the frequencies with charge and mass couplings on a fixed background. In the top row of panels we represent the real part of the frequency (or more precisely $1-\omega_R/\mu$). The thick curves correspond to the numerical results that we have generated through direct integration. We have requested a minimum relative precision of six digits (i.e. $10^{-6}$ of relative error). The thin solid lines (matching the color of the solid curves) correspond to the analytic guess, Eq.~\eqref{eq:ProcaOmegaR}. In the bottom panels we represent the imaginary part which is always negative. We also indicate with vertical lines the value where the  threshold condition  $\mu M=qQ$ is attained.

In the left panels we represent the fundamental mode frequency of the Proca system ($N=1$) with various curves for different positive and negative field charge as a function of mass coupling $\mu M$. The middle panels are similar except that we represent a first excited state ($N=2$). The main observations on these are:
\begin{itemize}
\item Generically, $\omega_I$ is negative and its modulus decreases for smaller mass. 
\item For negative field charge\footnote{Throughout we assume the convention that the BH charge is positive.} $q$, as it goes to more negative values, $|\omega_I|$ grows for fixed mass, which means that $\tau$ decreases so the scalar mode should be quickly absorbed by the BH.
\item For positive $q$ however, as we flow to smaller mass, there is a sharp threshold in $\omega_I$ precisely at $\mu M= qQ$, i.e. when the mass coupling balances out exactly the electromagnetic coupling. At this threshold $\tau\rightarrow +\infty$, which means that we can have arbitrarily long lived Proca quasi-bound states.
\item Furthermore, at the threshold $\mu M= qQ$, the real part of the frequency also tends sharply to $\mu$. This means that as we try to flow into the superradiant regime ($\omega<qQ$), we leave the quasi-bound state regime into the scattering regime, before reaching it. This explains the fact that we did not find any quasi-bound states in that regime.  
\item Finally, one can see a very good agreement between our guess, Eq.~\eqref{eq:ProcaOmegaR}, (thin solid lines) not only for small mass, but also near the long lived state threshold even for larger mass and charge couplings.
\end{itemize}
These features are confirmed in the right panels of Fig.~\ref{Fundamental}, where we have performed a flow of the charge $q$, with $Q$ close to the extremal BH background limit, for all the $N=1,2$ modes of the Proca system. Here it can be seen more clearly that as the charge $q$ is increased to larger positive values, there is a maximum value allowed which is given by $q_{max}=\mu M/Q$ (in the extremal limit this becomes $q=\mu$).

\begin{figure*}
\begin{center}
\includegraphics[clip=true, trim = 0 0 13 0, width=0.32\textwidth]{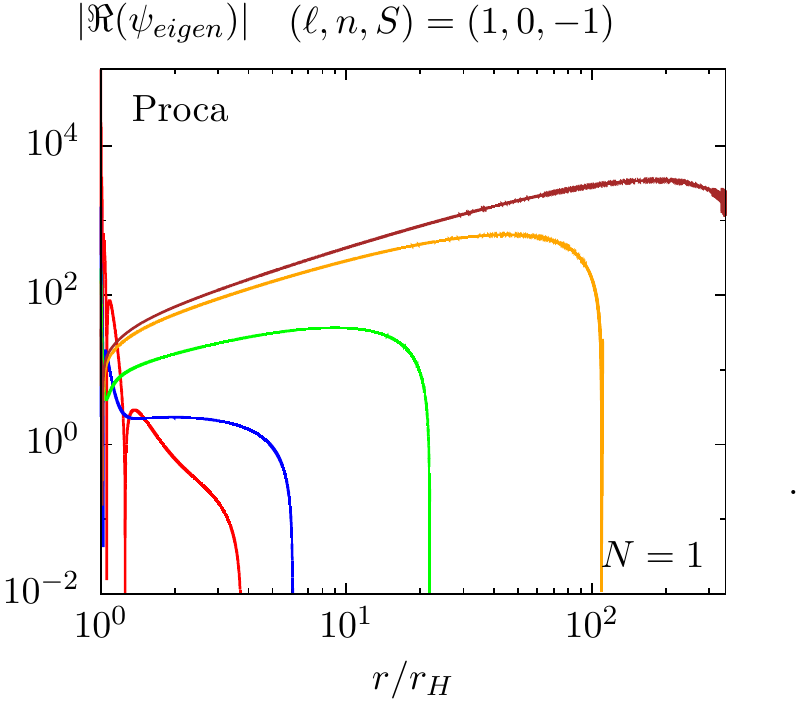}\hspace{1mm}\includegraphics[clip=true, width=0.32\textwidth]{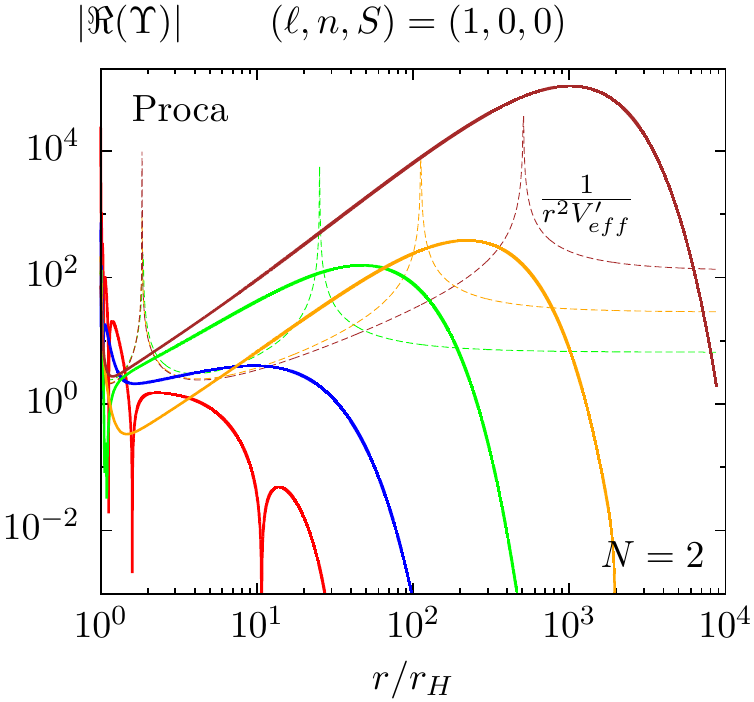}\hspace{1mm}\includegraphics[clip=true, width=0.32\textwidth]{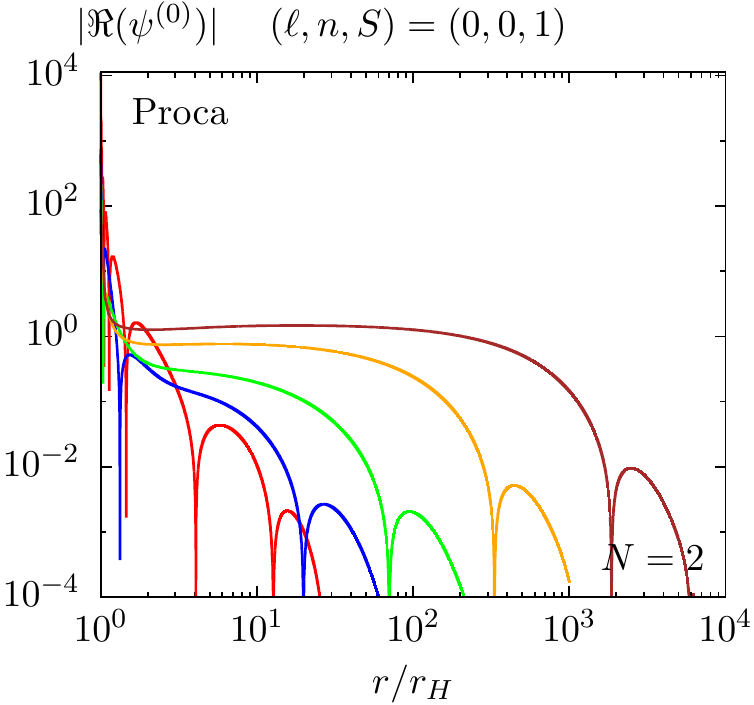} \vspace{2mm}\\
\includegraphics[clip=true, trim = 0 0 13 0, width=0.436\textwidth]{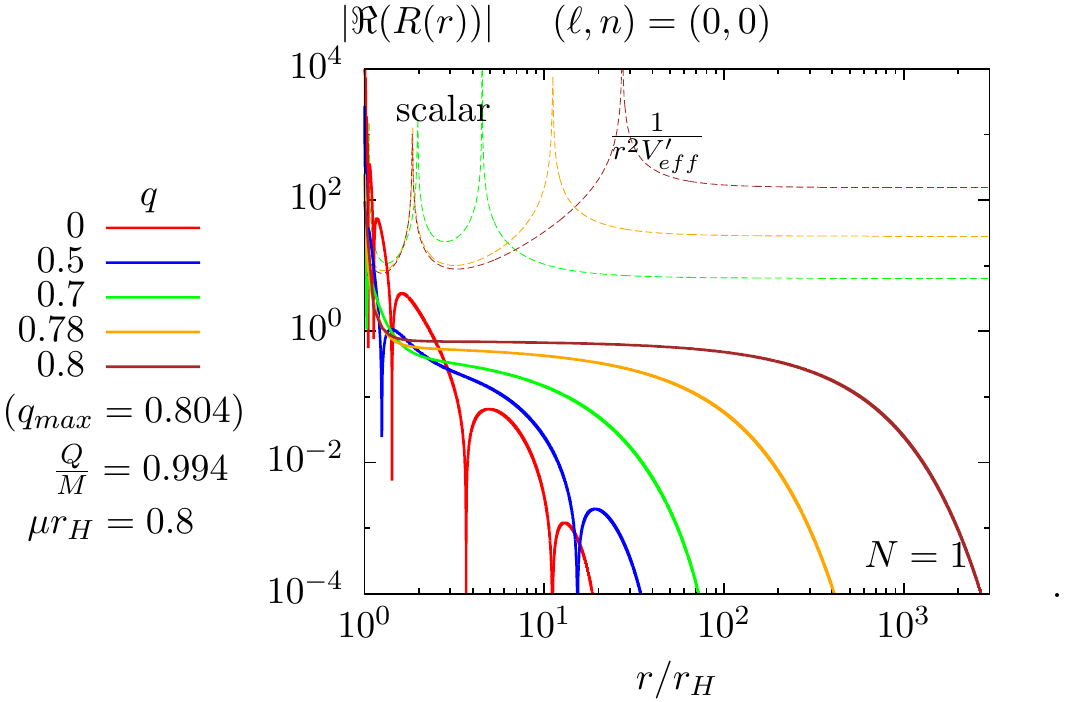}\hspace{3mm}\includegraphics[clip=true, trim = 0 0 13 0, width=0.35\textwidth]{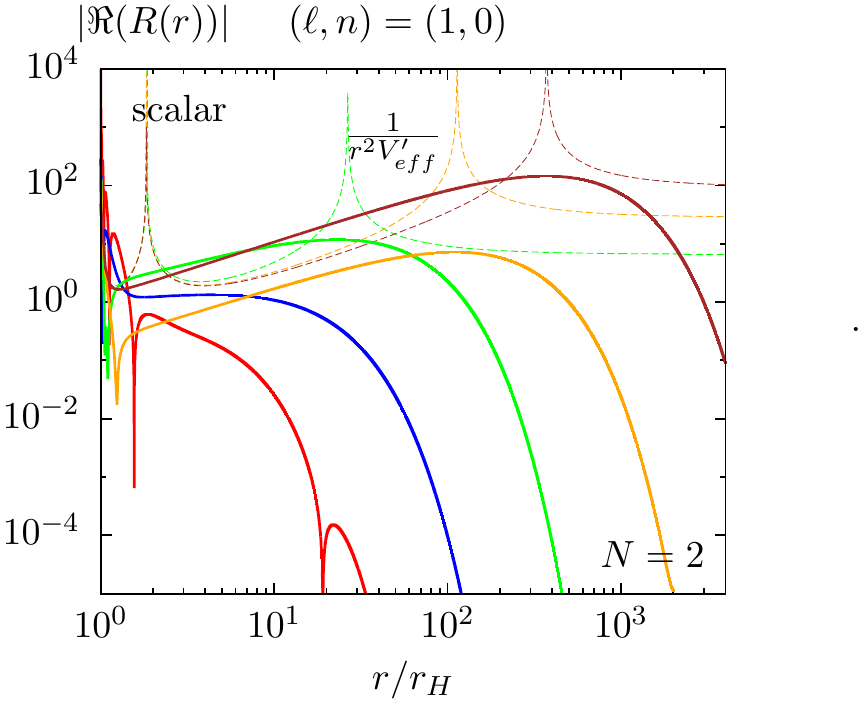}
\end{center}
\caption{\label{WFuncs} Radial profiles for some quasi-bound state modes for the Proca field (top row) and scalar field (bottom row) for comparison (key in the bottom row applies to all plots). We also indicate for some cases $(r^2dV_{eff}/dr)^{-1}$ whose poles indicate the local maximum and minimum of the effective potential (left and right peaks in each panel where it is represented). 
}
\end{figure*}

In Fig.~\ref{WFuncs} we display some examples of quasi-bound state radial profiles. In all panels we vary the field charge from zero up to a value close to the maximum allowed by the threshold condition, to observe how the field binds around the BH close to that limit. The top row is for modes of the Proca field and we have also included, for comparison, some modes for the scalar field in the bottom row. Whenever an effective potential is known, we display the quantity $(r^2dV_{eff}/dr)^{-1}$ (i.e. the inverse derivative). The latter has poles, one at a local maximum of the effective potential near the horizon and another at a minimum (further away), which define the location of the potential well.

In the top left panel we display the fundamental mode of the Proca field. This mode is in the coupled system and it is the eigenvector with zero eigenvalue of the $\mathbf{S}^{--}$ matrix. It results from the linear combination of two exponentially growing solutions, so for large $r$ the fine cancellation that must occur in the linear combination becomes numerically more difficult. This can be seen in the brown ($q=0.8$) curve which starts oscillating due to numerical errors on the right tail region (in $r$). It is clear from the curves that as we approach the threshold charge the field spreads out to a larger radius and the oscillations close to the horizon (which are associated with the field's absorption by the BH) become less extended. This shows that the field configuration becomes more and more bound state like (rather than quasi-bound). 

In the top middle panel we show an $N=2$ level obtained from the transverse $S=0$ equation. The features are similar to the left one except that the radial profile peaks more strongly. The peak is clearly correlated with the depth of the potential well where the field is more tightly bound, as seen in the dashed lines indicating the maximum and minimum of the potential.

The top right panel, shows the first $\ell=0$ level. The main difference is that the curves in the bound region tend to be much more flat. This is due to the absence of orbital angular momentum for these states. In fact, in the bottom left panel we show the corresponding mode in the scalar case (for comparison) where exactly the same happens for $\ell=0$.

In the scalar case (bottom row), similar features apply. In particular one can verify very clearly that the field profile in the long lived limit, extends inside the (increasingly wide) potential well.

\section{Analytic solutions}
\label{sec_analytic}
We will now show that in the limit in which the imaginary part of the frequency goes to zero, there are indeed non-trivial solutions, which are the limiting behavior of the modes found numerically in the previous section. This follows closely the analysis in~\cite{Degollado:2013eqa} but extends it, not only because we also consider also the Proca field but also because we do not restrict ourselves to the double extremal limit considered therein.

From the results of Sect.~\ref{results}, in the limit of the threshold condition
 \begin{equation}
\mu M=qQ\Leftrightarrow qQ=\dfrac{(1+Q^2)}{2}\mu
\label{forcebalance}
\end{equation}
the frequency becomes real and equal to the field mass: 
\begin{equation}
\omega = \mu \; .
\end{equation}
We replace these two conditions in the original equations, for both the Proca and the scalar field and define $\rho\equiv r-1$ and
\begin{equation}
\epsilon\equiv 1-Q^2 \ .
\end{equation}
If $\epsilon=0$, then $Q^2=1$; going back to natural units, this means $Q=M$ and $\mu=q$. So this corresponds to a \textit{double extremal limit}, i.e. an extremal limit for both the BH and the field. Thus $\epsilon$ parameterizes the deviation from the double extremal limit, but keeping the force balance condition (\ref{forcebalance}). We will show that there are exact solutions for the Proca field when $\epsilon=0$, an analogous situation to that seen in~\cite{Degollado:2013eqa} for the scalar field case. But even for $\epsilon\neq 0$ we can show the existence of non-trivial solutions. 

With the definitions above, the equations for the Proca field simplify as follows. The transverse mode equation becomes
\begin{eqnarray}
&&\left[\dfrac{d^2}{d\rho^2}+\left(\frac{1}{\rho} - \frac{2}{1+\rho} + \frac{1}{\epsilon+\rho}\right)\dfrac{d}{d\rho}-\dfrac{\ell(\ell+1)}{\rho(\epsilon+\rho)}+\right.  \label{TmodeEq2} \\
&&\phantom{................................................}\left.+\left(\dfrac{\epsilon\mu(1+\rho)}{2\rho(\epsilon+\rho)}\right)^2\right] \Upsilon=0  \nonumber \ .
\end{eqnarray}
The $\ell=0$ mode equation is in fact very similar
\begin{eqnarray}
&&\left[\dfrac{d^2}{d\rho^2}+\left(\frac{1}{\rho} + \frac{2}{1+\rho} + \frac{1}{\epsilon+\rho}\right)\dfrac{d}{d\rho}\right.  \label{Psi0_2} \\
&&\phantom{................}\left.-\dfrac{2-\epsilon}{\rho(1+\rho)(\epsilon+\rho)}+\left(\dfrac{\epsilon\mu(1+\rho)}{2\rho(\epsilon+\rho)}\right)^2\right] \psi^{(0)}=0  \nonumber \ .
\end{eqnarray}
The coupled system is now
\begin{eqnarray}
&&\left[\dfrac{d^2}{d\rho^2}+ \frac{2}{1+\rho}\dfrac{d}{d\rho}-\dfrac{\ell(\ell+1)}{\rho(\epsilon+\rho)}+\left(\dfrac{\epsilon\mu(1+\rho)}{2\rho(\epsilon+\rho)}\right)^2\right]\psi+\nonumber  \\ 
&&\phantom{...................}+i\mu\dfrac{4\rho^3+6\epsilon \rho^2+\epsilon^2(\rho-1)}{2\rho^2(\epsilon+\rho)^2}\chi=0 \ ; \nonumber \\
&&\left[\dfrac{d^2}{d\rho^2}-\dfrac{\ell(\ell+1)}{\rho(\epsilon+\rho)}+\left(\dfrac{\epsilon\mu(1+\rho)}{2\rho(\epsilon+\rho)}\right)^2\right]\chi-\dfrac{i\epsilon^2\mu(1+\rho)}{2\rho^2(\epsilon+\rho)^2}\psi=0.\nonumber \\ && \phantom{........................} \; \; \; \; \; \; \label{coupledequations2}
\end{eqnarray}
Finally, for the scalar field we have
\begin{eqnarray}
&&\left[\dfrac{d^2}{d\rho^2}+\left(\frac{1}{\rho} - \frac{2}{1+\rho} + \frac{1}{\epsilon+\rho}\right)\dfrac{d}{d\rho}-\dfrac{\ell(\ell+1)}{\rho(\epsilon+\rho)}+\right.  \label{RscalarEq2} \\
&&\phantom{..................}\left.+\left(\dfrac{\epsilon\mu(1+\rho)}{2\rho(\epsilon+\rho)}\right)^2+\dfrac{\epsilon(\rho-1)-2\rho}{\rho(1+\rho)^2(\epsilon+\rho)}\right] R=0  \nonumber \ .
\end{eqnarray}

Consider first the scalar field case and $\epsilon=0$. We recover the solution\footnote{The $\rho+1$ factor is due to the fact that we have included a $1/r$ in the scalar field ansatz.} in~\cite{Degollado:2013eqa}, $R=\sum_\ell R_\ell$, 
\begin{equation}
R_\ell=(\rho+1)\left(A^{(R)}\rho^{\ell}+B^{(R)}\rho^{-\ell-1}\right)\; .
\end{equation}
If we choose $A^{(R)}=0$, we get non-trivial solutions which are regular when $\rho\rightarrow+\infty$. Moreover, as shown in~\cite{Degollado:2013eqa}, even if each partial wave diverges at the horizon, appropriate combinations of them are \textit{regular at the horizon} and describe scalar particles at some spatial location in equilibrium with the BH (due to a force cancellation).

Considering the Proca field equations and $\epsilon=0$, we find exact general solutions which are qualitatively similar to the scalar solution:
\begin{eqnarray}
\Upsilon&=&((\rho+1)\ell+1)A^{(\Upsilon)}\rho^{\ell}+((\rho+1)\ell+\rho)B^{(\Upsilon)}\rho^{-\ell-1} \nonumber\\
\psi^{(0)}&=&\dfrac{1}{1+\rho}\left(A^{(0)}\rho+B^{(0)}\rho^{-2}\right) \nonumber\\
\chi&=& A^{(\chi)}\rho^{\ell+1}+B^{(\chi)}\rho^{-\ell} \nonumber\\
\psi&=& \dfrac{1}{1+\rho}\left[A^{(\psi)}\rho^{\ell+1}+B^{(\psi)}\rho^{-\ell}+\phantom{\dfrac{A}{B}}\right. \label{extremalsol}\\
&&\left.-i\mu\left(A^{(\chi)}\dfrac{\rho^{\ell+2}(3+\rho+\ell(2+\rho))}{(3+5\ell+2\ell^2)}-\right.\right. \nonumber\\
&& \phantom{.....................}\left.\left.-B^{(\chi)}\dfrac{\rho^{1-\ell}(\ell(2+\rho)-1)}{\ell(2\ell-1)}\right)\right] \ .\nonumber
\end{eqnarray}
One can re-analyze this limit working in isotropic coordinates following~\cite{Degollado:2013eqa} such that the line element and gauge field are
\begin{equation}
ds^2=-H^{-2}dt^2+H^2\delta_{ij}dx^idx^j\; , \; A=(H^{-1}-1)dt \ .
\end{equation}
$H(x^i)$ is a harmonic function on Euclidean 3-space, i.e. $\Delta_{\mathbb{E}^3}H=0$. To describe a single extremal RN BH, $H$ is taken to have a single pole $H=1+M/|{\bf x}|$. But multi-centered solutions, of Majumdar-Papapetrou type~\cite{Majumdar:1947eu,Papapetrou1945}, are also possible if one chooses a harmonic function with multiple poles.

 We now apply a scalar-vector decomposition, using invariant tensors on the Euclidean 3-space, of the form
\begin{eqnarray}
W_0&=&H^{-1}\Psi \\
\vec{W}&=&\nabla\Phi +\vec{V} \;\;, \;\; \nabla\cdot \vec{V}=0 \ ,
\end{eqnarray} 
where $W_\mu=(W_0,\vec{W})$.  All vectors and differential operators are now in Euclidean 3-space.
Then, $\Phi$ becomes the non-dynamical mode which disappears due to the (gauged) Lorentz condition; this condition is a consequence of the Proca equations. As such, we have (as before) a longitudinal degree of freedom, $\Psi$, and two transverse degrees of freedom, $\vec{V}$. From the time and space components of the Proca equation, these obey a Laplace and a Maxwell-Faraday like equation, respectively:
\begin{eqnarray}
\Delta \Psi&=&0 \\
\nabla\times \vec{\beta}&=&-i\mu \nabla\Psi 
\end{eqnarray}
where we have defined
\begin{equation}
\vec{\beta}=\dfrac{1}{H^2}\nabla \times \vec{V} \; .
\end{equation}
Clearly, the longitudinal mode equation can be solved by a point like source centered at any point
\begin{equation}
\Psi=\dfrac{A}{|\mathbf{x}-\mathbf{x}_0|}\; ,
\label{pc}
\end{equation}
 and it serves as a source to the transverse modes. Furthermore, due to the linear structure, we can superpose any linear combination of solutions. 
 
 The bottom line we wish to emphasize is that the marginal (charged) Proca clouds exist in the double extremal limit and, as the scalar clouds, can  be made regular at both the horizon and at infinity by choosing all $A^{(i)}=0$ in (\ref{extremalsol}) and by taking appropriate linear combinations of the remaining partial waves, as to obtain a solution of the form (\ref{pc}) with ${\bf x_0}$ not coinciding with the BH horizon.

When $\epsilon \neq 0$ we cannot find a closed form solution. Nevertheless, after asymptotically expanding the equations in the limit $\rho \gg 1$ we find 
\begin{eqnarray}
\Upsilon&\rightarrow &\rho\left(A^{(\Upsilon)}\rho^{J_{(0)}}+B^{(\Upsilon)}\rho^{-1-J_{(0)}}\right) \nonumber\\
\psi^{(0)}&\rightarrow&\dfrac{1}{\rho}\left(A^{(0)}\rho^{J_{(+)}}+B^{(0)}\rho^{-J_{(+)}-1}\right) \\
\chi&\rightarrow& A_1\rho^{J_{(-)}}+B_1\rho^{-1-J_{(-)}}+A_2\rho^{J_{(+)}}+B_2\rho^{-1-J_{(+)}}\nonumber \\
\rho\,\psi&\rightarrow&\hat{A}_1\rho^{2+J_{(-)}}+\hat{B}_1\rho^{1-J_{(-)}}+\hat{A}_2\rho^{2+J_{(+)}}+\hat{B}_2\rho^{1-J_{(+)}}\nonumber
\end{eqnarray}
where the hatted constants $\hat{A}_i,\hat{B}_i$ are proportional to $A_i,B_i$ (the constants are unimportant) and we have defined the effective total angular momentum
\begin{equation}
J_{(S)}\equiv \sqrt{\left(\frac{1}{2}+\ell+S\right)^2-\left(\frac{\epsilon \mu}{2}\right)^2}-\frac{1}{2} \; .
\end{equation}
 For the scalar field
\begin{equation}
R\rightarrow \rho\left(A^{(R)}\rho^{J_{(0)}}+B^{(R)}\rho^{-1-J_{(0)}}\right)\; .
\end{equation}
This result, justifies the existence of non-trivial solutions when the threshold condition $\mu M=qQ$ is obeyed,  even away from double extremality.

\section{Discussion}
\label{Discussion}

In this paper we have studied quasi-bound state test field configurations of charged Proca and scalar fields in the background of the RN BH. We have found some interesting properties which are present for both types of fields, which are likely to be generic for massive charged bosonic fields. 

Firstly, despite the existence of superradiant scattering of charged Proca fields by RN BHs~\cite{HSW:2012,Wang:2012tk}, we did not find any quasi-bound states in the superradiant regime, hence no superradiant instabilities due to charged Proca fields, in agreement with what occurs for charged scalar fields~\cite{Furuhashi:2004jk,Hod:2013nn,Hod:2013eea}. 

Secondly, we have found, considering $\mu M> q Q >0$, that either decreasing the mass $\mu$  or increasing the charge $q$ as to achieve the threshold condition  $\mu M=qQ$, one gets arbitrarily long lived states, since the imaginary part of the quasi-bound state frequency tends to zero. When taking this limit the frequency becomes equal to the field mass $\omega\rightarrow \mu$, and so the states become marginally bound. Since they do not trivialize, we obtain configurations we have dubbed \textit{marginal (charged) scalar and Proca clouds} around RN BHs. But observe that these clouds are qualitatively different from those found in the Kerr case~\cite{Hod:2012px,Hod:2013zza,Herdeiro:2014goa}, and recently extended to the Kerr-Newman case~\cite{Hod:2014baa}, which are real bound states. 

By analyzing the behavior of the field profiles when approaching the threshold condition, the region where the field has support agrees well with the region where a potential well is present. The width of the well typically increases when approaching the marginally bound limit meaning that the field has support on a wider region away from the BH horizon. Since the quasi-bound state is localized away from the horizon, the gravitational and electromagnetic field interactions which are responsible for supporting the bound state should be dominated by their asymptotic Newtonian and Coulombian limit respectively. Thus the threshold  condition $\mu M=qQ$ should in fact correspond to a force balance condition between the Newtonian force and the electrostatic force.

Scalar clouds around Kerr BHs can be promoted to non-linear hair. Indeed Kerr BHs with scalar hair exist that connect precisely to the Kerr solutions that allow the existence of the corresponding cloud~\cite{Herdeiro:2014goa,Herdeiro:2014ima,Herdeiro:2014jaa}. One may therefore ask if the existence of these marginal (charged) scalar and Proca clouds hints at the existence of non-linear solutions of RN BHs with scalar or Proca hair. One possibility is along the lines of the discussion in~\cite{Degollado:2013eqa}, for the scalar case. It was observed therein that the marginal scalar clouds (albeit this terminology was not used therein) can be seen as the partial waves of a distribution of charged scalar particles in a no-force balance with the RN BH. This suggests the existence of multi-object non-linear solutions -- of Majumdar-Papapetrou type~\cite{Majumdar:1947eu,Papapetrou1945}, but not necessarily multiple BHs. One other possibility, however, is that since the clouds that are regular at the horizon are a linear combination of partial waves with $\ell\neq 0$, the corresponding non-linear configurations will have non-zero angular momentum and thus will be a Kerr-Newman BH with charged scalar (or Proca) hair, a possibility already anticipated in~\cite{Herdeiro:2014goa,Hod:2014baa}.  Of course, the existence (or nonexistence) of any of these possible solutions, can only be decided by a fully non-linear analysis of the Einstein-(charged)-Klein-Gordon or Einstein-(charged)-Proca systems.

\bigskip

\noindent{\bf{\em Acknowledgements.}}
We would like to thank J. C. Degollado and E. Radu for many interesting discussions. M.W. and M.S. are funded by FCT through the grants SFRH/BD/51648/2011 and SFRH/BPD/69971/2010. The work in this paper is also supported by the grants PTDC/FIS/116625/2010 and  NRHEP--295189-FP7-PEOPLE-2011-IRSES.

\bibliographystyle{h-physrev4}
\bibliography{proca-charged}

\end{document}